\tolerance=10000
\magnification=1200
\baselineskip=20 pt

\centerline{{\bf CONSISTENT PERTURBATIVE LIGHT-FRONT}}
\centerline{{\bf FORMULATION OF QUANTUM ELECTRODYNAMICS}} 
\bigskip
\centerline{{\bf Michele Morara}}
\centerline{{\sl Dipartimento di Fisica "A. Righi",
                Universit\`a di Bologna,}}

\medskip

\centerline{and}

\medskip

\centerline{{\bf Roberto Soldati}
\footnote{$^\flat$}{E-mail: soldati@bo.infn.it}}
\centerline{{\sl Dipartimento di Fisica "A. Righi", Universit\'a
di Bologna}}
\centerline{{\sl and Istituto Nazionale di Fisica Nucleare, 
Sezione di Bologna,}}
\centerline{{\sl 40126 Bologna - Italy}}

\bigskip
\noindent
\centerline{Abstract}
\bigskip
\noindent
{\it A new light-front formulation of Q.E.D. is developed,
within the framework of standard perturbation theory,
in which $x^+$ plays the role of the evolution parameter and the gauge choice
is $A_+=0$ (light-front "temporal" gauge). It is shown that this formulation 
leads to the Mandelstam-Leibbrandt causal prescription for the non-covariant 
singularities in the photon propagator. Furthermore, it is proved that 
the dimensionally regularized one loop off-shell amplitudes exactly coincide 
with the correct ones, as
computed within the standard approach using ordinary space-time coordinates.}
\medskip
\noindent
PACS numbers : 11.10.Ef, 11.15.Bt, 12.20.-m
\bigskip
\noindent
DFUB-97-17
\par\noindent
April 98
\medskip
\centerline{{\sl Resubmitted to Phys. Rev. D}}

\vfill\eject
\noindent
{\bf 1.\ Introduction.} 
\medskip
Light-front formulation of gauge quantum field theories has become more 
and more popular in the last few years. In the abelian case, {\it i.e.}
standard Q.E.D. or abelian Higgs model, the renewal of the interest in 
that subject is mainly because of two reasons. 
On the one hand, the light-front hamiltonian
approach to Q.E.D. appears to provide an alternative tool to compute Lamb shift
[1] and deal with bound-state problems [2]. 
On the other hand, some non-perturbative
aspects - such as the role of the zero-modes [3] - have first to be clearly
understood in abelian models, before going into the much more
challenging non-abelian case. 

The original attempts to set up canonical quantization of Q.E.D. in the 
framework of light-front - or null-plane - dymamics date back to the early
seventies [4]. In the original approach, the light-cone coordinate
$x^+ = (x^0 +x^3)/\sqrt{2}$ plays the role of the evolution parameter and
the standard gauge choice is $A_-=0$, in such a way to stay as close as 
possible to the axial gauge formulation of Q.E.D. in standard space-time
coordinates (STC). 

After a considerable amount of work has been done along
this line, it was definitely discovered [5] that perturbation theory,
based upon the original light-front quantization scheme
for gauge theories, is inconsistent,
owing to loop integrations, just because the above scheme necessarily entails
the Cauchy principal value (CPV) prescription to understand the spurious
non-covariant poles in the gauge particle vector propagator. 
 This means that quite basic features of the
standard perturbative approach for gauge theories are lost, 
such as power counting
renormalizability, unitarity, covariance and causality. In other words,
the original approach to light-front quantization of gauge theories is certainly
not equivalent to the standard covariant formulation, already at the
perturbative level; it is {\it a fortiori} hard to believe that the same
approach could provide useful hints beyond perturbation theory, in the
absence of deep modifications. {\it En passant}, it is really curious 
and rather surprising that a non-negligeable fraction of the field
theorists community seems to have nowadays not yet fully gathered
and appreciated this rough breakdown of the conventional old light-front
approach to gauge theories. For instance, even the one loop Q.C.D.
beta function does not result to be, within that context,
the correct covariant one [5].

It has been noticed some time ago [6] that, in order to restore at least
causality for the free propagator of the gauge fields in the light-cone
gauge, a special prescription, thereof called the Mandelstam-Leibbrandt
(ML) prescription, has to be employed, in order to regulate the spurious
non-covariant singularities. Shortly afterwards, it has been realized
that the ML prescription arises from the canonical quantization in
standard STC, provided some special unphysical (ghost-like) degrees of
freedom are taken into account [7]. Even more, it has been proved that,
within that framework, gauge theories in the light-cone gauge are
renormalizable, unitary and covariant order-by-order in perturbation
theory [8]. It is worthwhile to emphasize how this remarkable result
crucially stems from the presence of the above mentioned unphysical
degrees of freedom: as soon as they are correctly taken into account,
the equivalence between the covariant and light-cone gauges is
established, within the standard perturbative approach in STC.

The open issue, which is still there, is to find a light-front formulation
for quantum gauge theories, which turns out to be equivalent to the 
conventional one in ordinary STC, at least in perturbation theory.
It is definitely clear, from the above considerations, that such a new
formulation, whatever it is, must lead to the ML prescription for the
non-covariant singularities of the gauge particle vector propagator,
at variance with the original old one, driving instead to the 
pathological CPV prescription.

A first step towards this direction has been done quite recently by
McCartor and Robertson [9]. They have found an algebraic scheme
to quantize the theory on the light-front, taking also the above mentioned
unphysical degrees of freedom into account. However, as they use the
"temporal" LCC as the evolution parameter and the "spatial" gauge choice
$A_- = 0$, the above algebraic setting is done after quantization
of physical and unphysical degrees of freedom on different 
characteristic surfaces, {\it i.e.} light-front hyperplanes. 
Beside being somewhat unnatural
\footnote{*}{Actually, in the presence of interaction, the  simultaneous
occurrence of "spatial" and "temporal" light-front hyperplanes, to
specify the operator's algebra, makes the treatment somewhat complicated.}
, this approach does not
drive exactly to the standard form of the photon propagator 
with the ML prescription for the spurious singularity. 
It is one of the aims of the present paper to show how the
latter drawbacks in the McCartor and Robertson approach could be
indeed overcome, without spoiling its correct content of an
enlarged light-front operator algebra.

In order to achieve this goal, we simply make the transition from the
"spatial" light-cone gauge $A_-=0$ to the "temporal" light-cone gauge
$A_+=0$, the "temporal" LCC $x^+$ being kept as the evolution parameter
within the light-front formulation. In so doing, on the one hand the
free field operator algebra for the whole set of fields is naturally
defined on the "spatial" hyperplanes $x^+ = {\rm constant}$. On the other
hand, the ML prescription is exactly recovered for the propagator of 
the free radiation field.

These remarkable features allow therefore to correctly develop perturbation
theory, once the corresponding 
interaction hamiltonian has been single out from
constraints analysis of (pseudo-)classical
Q.E.D. in LCC, including unphysical degrees of 
freedom ({\it i.e.} in an enlarged phase space). This leads to obtain
the set of light-front Q.E.D. Feynman's rules, 
which will be shown to involve an infinite set of special
non-covariant vertices. It is then amusing to 
check, at one loop, that truncated light-front Green's functions -
{\it i.e.} vacuum expectation values of light-front-time ordered
product of field operators - are exactly the same as in the usual
STC formulation, provided the gauge invariant dimensional regularization
scheme is embodied.

The paper is organized as follows. In Sect.2, we give a critical 
reading of the McCartor and Robertson approach to light-front quantization
of the free radiation field. In so doing, we point out where this approach
reveals to be unsatisfactory and how to implement it, in order to 
reproduce the ML form of the free propagator. In Sect.3 we briefly review
the light-front quantization of the free Dirac's field, in order to also
establish our notations for the light-front treatment of spinorial
matter. In Sect.4 we perform the canonical light-front quantization of  
Q.E.D. in the "temporal" light-cone gauge $A_+=0$, by means of the
standard Dirac's procedure for constrained systems. Sect.5 is devoted
to perturbation theory: namely, we derive Feynman's rules and show that,
up to the one loop approximation, dimensionally regularized
truncated and connected light-front Green's functions are the same,
as computed out of the standard canonical framework in usual STC.
Sect.6 contains some further comments and remarks, as well as an
outlook on future developments.

\bigskip
\noindent
{\bf 2.\ Light-front quantization of the free radiation field.} 
\medskip
Some time ago [7] it has been shown that the canonical quantization of
the free radiation field in the light-cone gauge $n^\mu A_\mu\equiv A_-=0,
\ (n^2=0)$,
is suitably formulated using standard space-time coordinates (STC) and leads,
eventually, to the ML prescription for the spurious singularities in the
propagator. It is worthwhile to stress that, in the derivation of the above 
result, the unphysical components of the gauge potential play a fundamental 
role. On the other hand, within the original approach to light-front
quantization using light-cone coordinates (LCC) [4], those unphysical 
degrees of freedom turn out to fulfil constraint equations instead of 
genuine equations of motion. 
Thereby, they are eliminated after imposing suitable boundary
conditions and, consequently,
only the physical degrees of freedom are indeed submitted to
canonical quantization. In so doing, unfortunately, the spurious singularity
in the vector propagator results to be prescribed as Cauchy principal valued
and leads to an inconsistent meaningless perturbation theory.

It is our aim to show in this section how some light-front quantization scheme
exists for the radiation field in LCC, which drives eventually to the ML
prescrition for the spurious singularity of the vector propagator, just
like the standard STC formulation does. In order to achieve this goal, 
we will develop and improve a recent attempt [9], in which the above
mentioned unphysical components of the gauge potential are retained and
quantized in LCC according to a new procedure. Let us first briefly review 
the main points of this approach. 

The starting point is the lagrangean density of the free radiation field
$$
{\cal L}_{{\rm rad}}=-{1\over 4}F_{\mu\nu}F^{\mu\nu}-\Lambda n^\mu A_\mu\ ,
\eqno(2.1)
$$
where $n_\mu=(n_+,n_\perp, n_-)=(1,0,0,0)$, 
in such a way that $n^\mu A_\mu=A_-$, 
and $\Lambda$ is a Lagrange multiplier which enforces the gauge constraint.

The Euler-Lagrange equations lead to
$$
\eqalignno{
\partial_\mu F^{\mu\nu} & =n^\nu\Lambda\ , &(2.2a)\cr
A_- & =\ 0\ . &(2.2b)\cr}
$$
It is convenient to introduce some new field variables as follows: namely,
$$
\eqalignno{
& A_\alpha = T_\alpha -{\partial_\alpha\over \partial^2_\perp} \varphi\ ,
&(2.3a)\cr
& A_+ = {\partial_\alpha\over \partial_-} T_\alpha -
{\partial_+\over \partial^2_\perp} \varphi\ -
{1\over \partial^2_\perp}\Lambda\ ;
&(2.3b)\cr}
$$
then Eq.s~(2.2) become
$$
\eqalignno{
(2\partial_+\partial_- & -\partial_\perp^2)T_\alpha=0\ , &(2.4a)\cr
\partial_-\varphi & =\partial_-\Lambda=0\ . &(2.4b)\cr}
$$

We notice that, as the fields $T_\alpha (x)$ fulfil free D'Alembert's
equations of motion, then the inverse of the light-front-space derivative
in eq.~(2.3b) is understood here to be $(1/\partial_-)\equiv
(2\partial_+/\partial_\perp^2)$. Furthermore,
from eq.~(2.4b) we can easily see that the fields $\varphi$ and $\Lambda$
do not fulfil evolution equations - remember that here it is the LCC $x^+$
which plays the role of the evolution parameter - but, as previously noticed, 
they satisfy constraint equations and, therefrom, can not be canonically
quantized on the null hyperplanes at constant $x^+$.

Now, it has been suggested [9],~[10] a new light-front quantization 
procedure, in which the transverse fields $T_\alpha$ are quantized 
on null hyperplanes at
equal $x^+$, according to the original light-front recipe, while the
longitudinal fields  $\varphi$ and $\Lambda$ at equal $x^-$.
Following this procedure, one can set up the 
generators of the translations on the null hyperplanes $\Sigma_+$ and
$\Sigma_-$, in the limit $L\to\infty$ (see Fig.1), and obtain, taking  
the Heisenberg equations of motion (2.4)
into account, the commutation relations
$$
\eqalignno{
&\left[T_\alpha (x),T_\beta (y)\right]_{x^+=y^+}  =
-{i\over 2}\delta_{\alpha\beta}\delta^{(2)}
(x^\perp-y^\perp){\rm sgn} (x^--y^-)\ , &(2.5a)\cr
&\left[\varphi (x),\Lambda (y)\right]_{x^-=y^-}  =
i\delta (x^+ -y^+)\partial_\perp^2\delta^{(2)}
(x^\perp-y^\perp)\ , &(2.5b)\cr
&\left[T_\alpha (x),\varphi (y)\right] =\left[T_\alpha (x),\Lambda (y)\right]
=\left[\varphi(x),\varphi (y)\right]=\left[\Lambda (x),\Lambda (y)\right]=0\ ,
&(2.5c)\cr}
$$
where ${\rm sgn} (x)$ denotes the usual sign distribution.
In so doing, the Authors of Ref.s [9] suggest that the light-cone-time
ordered product of the gauge potential operators defined by
$$
D^+_{\mu\nu}(x-y)\equiv\theta (x^+ -y^+)\left<0|A_\mu (x) A_\nu (y)|0\right>
+\theta (y^+ -x^+)\left<0|A_\nu (y) A_\mu (x)|0\right>\ ,
\eqno(2.6)
$$
might eventually give rise to the ML form of the gauge field propagator.
Actually we shall show below that this is not exactly true, owing to the 
presence of some ill-defined products of tempered distributions.
\midinsert
\vskip 8 truecm
\includegraphics{figura1.eps}
\endinsert
As a matter of fact, if we consider the longitudinal components of the 
gauge potential: namely,
$$
\Gamma_\mu=-{1\over \partial_\perp^2}\left(\partial_\mu\varphi+
n_\mu\Lambda\right)\ ,
\eqno(2.7)
$$
then a straightforward calculation yields
$$
\left<0|\Gamma_\mu (x)\Gamma_\nu (y)|0\right>=\int{d^4 k\over (2\pi)^3}
e^{ik(x-y)}\theta (-k_+)\delta (k_-){n_\mu k_\nu + n_\nu k_\mu\over
k_\perp^2}\ .
\eqno(2.8)
$$
After multiplication, for instance, with $\theta (x^+ -y^+)$ and taking
the Fourier transform we formally get the convolution
$$
 \int_{-\infty}^0 {d\xi\over 2\pi i}{\delta (k_-)\over (k_+ -\xi)-i\epsilon}
\left[{n_\mu k_\nu + n_\nu k_\mu\over
k_\perp^2}\right]_{k_+=\xi}\ . 
$$
One can easily convince himself that the above expression does not define
a tempered distribution - owing to the logarithmic divergence in the
$\xi$-integration - which means, in turn, that the propagator in eq.~(2.6)
is not properly understood from the mathematical point of view.

Nonetheless, it is indeed remarkable that the main idea behind the
quantization procedure in Ref.s [9], {\it i.e.} the enlarged algebra
on the characteristic surfaces in order to fulfil causality,
is suggestive, albeit troubles arise when dealing with the 
evolution. It should be apparent that, in fact, the very same reasons 
preventing us from specifying the algebra of the longitudinal field
operators at equal $x^+$, also prevent us from propagating the
unphysical degrees of freedom along $x^+$. The simplest way to circumvent 
these difficulties and to build up a consistent light-front dynamics turns
out to be a change of the null gauge vector\footnote{$^\sharp$}{Actually, an
equivalent way to proceed is to keep the previous light-cone gauge
choice unaltered and to change the evolution parameter (the light-front-time)
from $x^+$ to $x^-$, the key point being that the light-front-time and the
light-cone gauge vector have to be parallel.}, {\it i.e.} we replace
$n_\mu\longmapsto n^*_\mu\equiv (0,0,0,1)$ in such a way that $n^{*\mu}A_\mu=
A_+=0$.

Let us therefore consider the new lagrangean density
$$
{\cal L}_{{\rm rad}}=-{1\over 4}F_{\mu\nu}F^{\mu\nu}-\Lambda n^{*\mu} A_\mu\ ;
\eqno(2.9)
$$
as the whole set of fields now fulfils genuine equations of motion, it is 
convenient to proceed within the framework of Dirac's canonical quantization
[11].

The canonical momenta are ( ${\cal A}_{{\rm rad}}\equiv 
\int d^4 x {\cal L}_{{\rm rad}}(x)$ )
$$
\eqalignno{
& \pi^-  \equiv {\delta {\cal A}_{{\rm rad}}\over \delta\partial_+ A_-} 
        = F_{+-}\ , & (2.10a)\cr
& \pi^\alpha \equiv {\delta {\cal A}_{{\rm rad}}\over \delta\partial_+ 
A_\alpha}
        = F_{-\alpha}\ , & (2.10b)\cr
& \pi^+\equiv {\delta {\cal A}_{{\rm rad}}\over \delta\partial_+ A_+}
        = 0\ , & (2.10c)\cr
& \pi^\Lambda\equiv {\delta {\cal A}_{{\rm rad}}\over \delta\partial_+ 
\Lambda}
              = 0\ , & (2.10d)\cr}
$$
whence it follows that there are two primary second class constraints (2.10b)
originating from the use of LCC, as well as two primary first class constraints
(2.10c-d).

The canonical hamiltonian becomes
$$
H_{{\rm rad}}=\int d^3 x\left\{{1\over 2}\left(\pi^-\right)^2+{1\over 4}
F_{\alpha\beta}F_{\alpha\beta} - A_+\left(\partial_\alpha\pi^\alpha +
\partial_-\pi^- - \Lambda\right)\right\}\ ,
\eqno(2.11)
$$
and, consequently, from the light-front-temporal consistency of the first class
constraints (2.10c-d) we derive the secondary constraints
$$
A_+=0\ ,
\eqno (2.12a)
$$
$$
\partial_\alpha\pi^\alpha +\partial_-\pi^- - \Lambda = 0\ .
\eqno(2.12b)
$$

The full set of constraits is now second class and thereby we can compute the
Dirac's brackets. After choosing as independent fields the following ones,
$$
\phi_1=A_1\ ,\ \phi_2=A_2\ ,\ \phi_3=A_-\ ,\ \phi_4=\pi^-\ ,
\eqno(2.13)
$$
we eventually obtain the Dirac's brackets matrix 
$$
\Phi_{ab}({\bf x},{\bf y})\equiv
\left.\left\{\phi_a (x),\phi_b (y)\right\}_D\right|_{x^+=y^+}\ ,\quad
a,b=1,2,3,4\ ,
$$
whose matrix elements are integro-differential operators 
in terms of light-front-space coordinates ${\bf x}= (x^1,x^2,x^-)$: namely,
$$
\Phi_{ab}({\bf x},{\bf y})=\left|
\matrix{-{\bf 1}/2\partial_- & 0 & 0 & \partial_1/2\partial_- \cr
       0 & -{\bf 1}/2\partial_- & 0 & \partial_2/2\partial_- \cr
       0 & 0 & 0 & {\bf 1} \cr
       -\partial_1/2\partial_- & -\partial_2/2\partial_- &
       -{\bf 1} & \partial_\perp^2/2\partial_- \cr}\right|\ .
\eqno(2.14)
$$
Here the identity ${\bf 1}$ means the product $\delta (x^- -y^-)\delta^{(2)}
(x^\perp-y^\perp)$, whilst the kernels $({\bf 1}/\partial_-)$
and $(\partial_\alpha/\partial_-)$ are shorthands for
$\delta^{(2)} (x^\perp-y^\perp){\rm sgn}(x^- -y^-)$ and
$\left(\partial_\alpha\delta^{(2)}\right) 
(x^\perp-y^\perp){\rm sgn}(x^- -y^-)$ respectively.
It should be noticed that the sign distribution is such 
to enforce standard (anti-)symmetry properties of 
Dirac's brackets.

After setting the secondary constraints strongly equal to zero in the 
hamiltonian (2.11), we obtain the Dirac's form
$$
H_D=\int d^3 x\left\{{1\over 2}\left(\pi^-\right)^2+{1\over 4}
F_{\alpha\beta}F_{\alpha\beta}\right\}\ .
\eqno(2.15)
$$
Now, in order to simplify the equations of motion, it is convenient to make the
change of variables similar to the one of Eq.s~(2.3) but taylored to the present
light-cone gauge choice $A_+=0$: namely,
$$
\eqalignno{
& A_\alpha = T_\alpha -{\partial_\alpha\over \partial^2_\perp}\varphi\ ,
&(2.16a)\cr
& A_- = {2\partial_-\over \partial_\perp^2}\partial_\alpha T_\alpha -
{\partial_-\over \partial^2_\perp}\varphi\ -
{1\over \partial^2_\perp}\Lambda\ ;
&(2.16b)\cr
& \pi^-=\partial_\alpha T_\alpha\ . &(2.16c)\cr}
$$

The Dirac's brackets among the new independent fields read
$$
\Phi^\prime_{ab}({\bf x},{\bf y})=\left|
\matrix{-{\bf 1}/2\partial_- & 0 & 0 & 0 \cr
       0 & -{\bf 1}/2\partial_- & 0 & 0 \cr
       0 & 0 & 0 & \partial_\perp^2 \cr
       0 & 0 & -\partial_\perp^2 & 0 \cr}\right|\ .
\eqno(2.17)
$$
where we have set 
$$
\phi^\prime_1=T_1\ ,\ \phi^\prime_2=T_2\ ,\ \phi^\prime_3=\varphi\ ,\ 
\phi^\prime_4=\Lambda\ ,
\eqno(2.18)
$$
and the canonical hamiltonian takes its final Dirac's form
$$
H_D^\prime\equiv \int d^3 x\left\{{1\over 2}
\partial_\beta T_\alpha\partial_\beta T_\alpha\right\}\ ,
\eqno(2.19)
$$
whence we obtain the genuine equations of motion
$$
\eqalignno{
\partial_+ T_\alpha & = {\partial_\perp^2\over 2\partial_-}T_\alpha\ , 
&(2.20a)\cr
\partial_+\varphi & = 0\ , &(2.20b)\cr
\partial_+\Lambda & = 0\ . &(2.20c)\cr}
$$

The transition to the quantum theory is accomplished under replacement of the
Dirac's brackets with canonical equal light-front-time commutation relations,
which read
$$
\eqalignno{
&\left[T_\alpha (x),T_\beta (y)\right]_{x^+=y^+}  =
-{i\over 2}\delta_{\alpha\beta}\delta^{(2)}
(x^\perp-y^\perp){\rm sgn}(x^--y^-)\ , &(2.21a)\cr
&\left[\varphi (x),\Lambda (y)\right]_{x^+=y^+}  =
i\delta (x^- -y^-)\partial_\perp^2\delta^{(2)}
(x^\perp-y^\perp)\ , &(2.21b)\cr
&\left[T_\alpha (x),\varphi (y)\right] =\left[T_\alpha (x),\Lambda (y)\right]
=\left[\varphi(x),\varphi (y)\right]=\left[\Lambda (x),\Lambda (y)\right]=0\ .
&(2.21c)\cr}
$$
It is important to notice that the above canonical commutation relations (CCR) 
have the very same form as in the McCartor and Robertson 
quantization scheme, see Eq.s~(2.5),
up to the crucial difference that now the quantization 
characteristic surface is the same
for all the fields.

Let us now search for the solutions, in the framework of the tempered 
distributions, of the equations of motion in the Fourier space. To this
aim, it is convenient to introduce again
the longitudinal (unphysical) components
of the radiation field
$$
\Gamma_\mu = -{1\over \partial_\perp^2}\left(\partial_\mu\varphi + 
n^*_\mu\Lambda\right)\ ,
\eqno(2.22)
$$
in such a way that 
$$
T_\mu (x)\equiv A_\mu (x) - \Gamma_\mu (x)\ .
\eqno(2.23)
$$
For the transverse components we easily get
$$
T_\mu (x) =
\int {d^2 k_\perp dk_-\over (2\pi)^{3/2}}{\theta (k_-)\over \sqrt{2k_-}}
\varepsilon_\mu^\alpha (k_\perp ,k_-)\left\{a_\alpha (k_\perp ,k_-) e^{-ikx} +
a^\dagger_\alpha (k_\perp ,k_-)e^{ikx}\right\}_{k_+=k_\perp^2/2k_-}\ ,
\eqno(2.24)
$$
where the (real) polarization vectors are given by
$$
\varepsilon_\mu^1 (k_\perp ,k_-)=\left|
\matrix{0\cr 1\cr 0\cr 2k_1 k_-/k_\perp^2\cr}\right|\ ,\quad
\varepsilon_\mu^2 (k_\perp ,k_-)=\left|
\matrix{0\cr 0\cr 1\cr 2k_2 k_-/k_\perp^2\cr}\right|\ ,
\eqno(2.25)
$$
whilst the longitudinal components read
$$
\Gamma_\mu (x) =\int {d^2 k_\perp dk_-\over (2\pi)^{3/2}}
{\theta (k_-)\over \sqrt{k_\perp}}\left\{\left[
-{k_\mu\over k_\perp}f(k_\perp ,k_-)+n^*_\mu g(k_\perp ,k_-)\right]
e^{-ikx} +\ {\rm h.\ c.}\right\}_{k_+=0}\ ,
\eqno(2.26)
$$
where $k_\perp\equiv \sqrt{k_1^2+k_2^2}$.
The canonical commutation relations (2.21) entail the following algebra
of the creation-annihilation operators: namely,
$$
\eqalignno{
\left[a_\alpha (k_\perp ,k_-),a_\beta^\dagger (p_\perp ,p_-)\right] & =
\delta_{\alpha\beta}\delta^{(2)}(k_\perp -p_\perp)\delta (k_- -p_-)\ ,
& (2.27a)\cr
\left[f(k_\perp ,k_-),g^\dagger (p_\perp ,p_-)\right] & =
\delta^{(2)}(k_\perp -p_\perp)\delta (k_- -p_-)\ ,
& (2.27b)\cr
\left[g(k_\perp ,k_-),f^\dagger (p_\perp ,p_-)\right] & =
\delta^{(2)}(k_\perp -p_\perp)\delta (k_- -p_-)\ ,
& (2.27c)\cr}
$$
all the other commutators vanishing.

The canonical commutation relations (2.27b-c) show that the theory involves
an indefinite metric space of states. The physical subspace 
${\cal V}_{{\rm phys}}$, 
whose metric turns out to be positive semi-definite, is defined through the
condition [7]:
$$
g(k_\perp ,k_-)\left|{\bf v}\right> = 0\ ,\quad 
\forall\left|{\bf v}\right>\in {\cal V}_{{\rm phys}}\ .
\eqno(2.28)
$$
It should be noted that, as
$$
\left<{\bf w}|\Lambda (x)|{\bf v}\right>=0\ ,\quad
\forall\left|{\bf w}\right>,\left|{\bf v}\right>\in {\cal V}_{{\rm phys}}\ ,
\eqno(2.29)
$$
the Gauss' law is indeed fulfilled in ${\cal V}_{{\rm phys}}$.

Let us finally compute the free vector propagator
$$
D^+_{\mu\nu}(x-y)\equiv\theta (x^+ -y^+)\left<0|A_\mu (x) A_\nu (y)|0\right>
+\theta (y^+ -x^+)\left<0|A_\nu (y) A_\mu (x)|0\right>\ ,
\eqno(2.30)
$$
which, after the gauge fixing condition (2.12a), turns out to be properly
defined from the mathematical point of view, {\it i.e.} the product of the
distributions in eq.~(2.30) does indeed exist. Separating the transverse and 
longitudinal components, setting 
$a_{\mu\nu}(k)\equiv n^*_\mu k_\nu +n^*_\nu k_\mu$ and going to the momentum 
space we eventually get
$$
\tilde D^T_{\mu\nu} (k)={i\over k^2 + i\epsilon}\left[-g_{\mu\alpha}
g^\alpha_\nu + 
{2k_-\over k_\perp^2} a_{\mu\nu} (0,k_\perp ,k_-)\right]\ ,
\eqno(2.31)
$$
$$
\tilde D^{\Gamma}_{\mu\nu} (k)= -i{k_-\over k_+ k_- +i\epsilon}
{a_{\mu\nu} (0,k_\perp,k_-)\over k_\perp^2}\ .
\eqno(2.32)
$$
Taking into account that
$$
{2k_-\over k_\perp^2}
\left({1\over k^2+i\epsilon}-{1\over 2k_-k_+ +i\epsilon}\right)=
{1\over k^2+i\epsilon}{1\over [k_+]}\ ,
\eqno(2.33)
$$
where
$$
{1\over [k_+]}\equiv {1\over k_+ +i\epsilon
{\rm sgn}(k_-)}\equiv {k_-\over k_-k_+ +i\epsilon}\ ,
\eqno(2.34)
$$
which is nothing but the Mandelstam-Leibbrandt distribution, 
we finally get the propagator in the momentum space
$$
\tilde D^+_{\mu\nu} (k)={i\over k^2+i\epsilon}\left[
-g_{\mu\nu}+{n^*_\mu k_\nu + n^*_\nu k_\mu\over [n^*k]}\right]\ .
\eqno(2.35)
$$
It has to be stressed that, more than being mathematically well defined,
 the present form of the free vector propagator
exactly coincides with the one obtained in the framework of ordinary time
canonical quantization of Ref.~[7]. This means that the light-front
operator algebra (2.21) together with light-front-time propagation are
completely equivalent, at the level of the free field theory, to the 
ordinary time canonical quantization and standard chronological pairing,
at variance with the old light-front formulation of Ref.s~[4].
This non trivial result, which arises as the correct implementation of
the original ideas of Ref.s~[9], will survive after the switching on 
of the interaction with spinor matter, as we shall discuss below.
\bigskip
\noindent
{\bf 3.\ Light-front quantization of the free Dirac field.}
\bigskip
Before going to the treatment of Q.E.D. it is useful to briefly review the 
canonical
light-front quantization of the free Dirac field and, in so doing, establish 
our conventions and notations.
First we recall that, in order to obtain the correct canonical anticommutation
relations from Dirac's procedure, it is convenient to consider the system 
at the (pseudo)classical level. This means that we start from spinor fields 
in terms of Grassmann-valued fields satysfying the graded version of the
canonical Poisson's and Dirac's brackets (see, for instance Ref.~[12]).
The same formalism will be generalized in the next Section, where Bose fields
are also included.

Within the framework of the light-front quantization, it is customary to
introduce the following representation of the Dirac's matrices: namely,
$$
\gamma^+=\left|\matrix{0 & 0\cr \sqrt{2}\sigma^1 & 0\cr}\right|\ 
\gamma^1=\left|\matrix{-i\sigma^2 & 0\cr 0 & -i\sigma^2\cr}\right|\ 
\gamma^2=\left|\matrix{i\sigma^1 & 0\cr 0 & -i\sigma^1\cr}\right|\ 
\gamma^-=\left|\matrix{0 & \sqrt{2}\sigma^1\cr 0 & 0}\right|\ ,
\eqno(3.1)
$$
and we write the four-component Dirac's spinor as
$$
\Psi\equiv \left|\matrix{\psi\cr \chi\cr}\right|\ ,
\eqno(3.2)
$$
with $\psi ,\chi$ two-components complex spinors. Here $\sigma^i,\ i=1,2,3$
are the Pauli's matrices and we also set
$$
\tau^1\equiv \sigma^3\ ,\quad \tau^2 \equiv i{\bf 1}_2\ .
\eqno(3.3)
$$
Therefore, the lagrangean density for the free Dirac's field
$$
{\cal L}_D = \bar \Psi\left(i\gamma^\mu\partial_\mu - m\right)\Psi\ ,
\eqno(3.4)
$$
where $\bar \Psi\equiv \Psi^\dagger\gamma^0 ,\ \gamma^0=2^{-1/2}(\gamma^+
+\gamma^-)$, may be rewritten as
$$
{\cal L}_D=\psi^\dagger i\sqrt{2}\partial_+\psi +
\chi^\dagger i\sqrt{2}\partial_-\chi +
\psi^\dagger\left(i\tau^{\alpha\dagger}\partial_\alpha -m\sigma^1\right)\chi
 +\chi^\dagger\left(i\tau^{\alpha}\partial_\alpha -m\sigma^1\right)\psi\ ,
\eqno(3.5)
$$
whence the canonical momenta read
$$
\eqalignno{
\pi^\psi & = -i\sqrt{2}\psi^\dagger\ , & (3.6a)\cr
\pi^{\psi^\dagger} & = 0\ , & (3.6b)\cr
\pi^{\chi} & = 0\ , & (3.6c)\cr
\pi^{\chi^\dagger} & = 0\ . & (3.6d)\cr}
$$

It follows that we have two primary second class constraints (3.6a-b) and
two primary first class constraints (3.6c-d). The canonical hamiltonian
turns out to be
$$
H=\int d^3 x\left\{
-\chi^\dagger i\sqrt{2}\partial_-\chi -
\psi^\dagger\left(i\tau^{\alpha\dagger}\partial_\alpha -m\sigma^1\right)\chi
 -\chi^\dagger\left(i\tau^{\alpha}\partial_\alpha -m\sigma^1\right)\psi
\right\}
\eqno(3.7)
$$
and the light-front-temporal consistency of the first class constraints
lead to the onset of the secondary constraints
$$
\eqalignno{
& i\sqrt{2}\partial_-\chi^\dagger + i\partial_\alpha\psi^\dagger
\tau^{\alpha\dagger} + m\psi^\dagger\sigma^1 = 0\ , & (3.8a)\cr
& i\sqrt{2}\partial_-\chi +\left(i\tau^\alpha\partial_\alpha 
- m\sigma^1\right)\psi = 0\ . & (3.8b)\cr}
$$

Now, the whole set of constraints being second class, the graded
Dirac's bracket can be consistently defined and taking $\psi$ and
$ \psi^\dagger$
as independent fields we readily find
$$
\left.\left\{\psi_r (x),\psi^\dagger_{r^\prime}(y)\right\}_D
\right|_{x^+=y^+}=
{1\over \sqrt2}\delta_{rr^\prime}\delta^{(2)}(x^\perp -y^\perp)
\delta(x^- -y^-)\ ,\quad r,r^\prime=1,2\ ,
\eqno(3.9)
$$
all the other graded Dirac's brackets vanishing.

After solving the secondary constraints (3.8) in terms of the independent
fields $\psi ,\ \psi^\dagger$ the canonical hamiltonian (3.7) can be cast 
into Dirac's form: namely,
$$
H_D =i\sqrt2\int d^3 x\left\{\psi^\dagger{\partial_\perp^2 - m^2\over
2\partial_-}\psi\right\}\ ,
\eqno(3.10)
$$
from which we obtain the canonical equations of motion
$$
\eqalignno{
\partial_+\psi_r & = {\partial_\perp^2 - m^2\over 2\partial_-}\psi_r\ ,
& (3.11a)\cr
\partial_+\psi^\dagger_r & = 
{\partial_\perp^2 - m^2\over 2\partial_-}\psi^\dagger_r\ ,
& (3.11a)\cr}
$$
showing that the independent fields $\psi ,\ \psi^\dagger$ correctly
fulfil the Klein-Gordon equation.

The expansion into normal modes leads to the standard decomposition
$$
\eqalign{
\psi (x) & =\int {d^3 k\over (2\pi)^{3/2}}{\theta (k_-)\over 2^{1/4}}\cr
& \times  
\sum_{s=\pm 1/2}\left\{w^s b_s (k_\perp ,k_-)e^{-ikx}
+w^{-s} d^\dagger_s (k_\perp ,k_-)
e^{ikx}\right\}_{k_+=(k_\perp^2 +m^2)/2k_-}\ ,\cr}
\eqno(3.12)
$$
and hermitean conjugate, where the polarization vectors are simply given by
$$
w^{s=1/2}\equiv \left|\matrix{1\cr 0\cr}\right|\ ,\quad
w^{s=-1/2}\equiv \left|\matrix{0\cr 1\cr}\right|\ .
\eqno(3.13)
$$ 
 
As it is well known the graded Dirac's brackets (3.9) entail the 
canonical operator algebra
$$
\eqalignno{
\left\{b_s (k_\perp ,k_-),b_{s^\prime}^\dagger (p_\perp ,p_-)\right\} & =
\delta_{ss^\prime}\delta^{(2)}(k_\perp -p_\perp)\delta (k_- -p_-)\ ,
& (3.14a)\cr
\left\{d_s (k_\perp ,k_-),d_{s^\prime}^\dagger (p_\perp ,p_-)\right\} & =
\delta_{ss^\prime}\delta^{(2)}(k_\perp -p_\perp)\delta (k_- -p_-)\ ,
& (3.14b)\cr}
$$
all the other anticommutators vanishing.

We are now ready to compute the free light-front fermion propagator
which is defined to be
$$
iS^+(x-y)\equiv
\theta (x^+-y^+)\left<0|\Psi (x)
\bar \Psi (y)|0\right> - 
\theta (y^+-x^+)\left<0|\bar \Psi (y)
\Psi (x)|0\right>\ .
\eqno(3.15)
$$
To this aim, it is convenient to introduce the light-front pairing between
two-components spinors $\alpha_r ,\ \beta_{r^\prime},\ r,r^\prime =1,2$, in 
such a way that
$$
S^{+\alpha\beta}_{rr^\prime}(x-y)\equiv 
\theta (x^+-y^+)\left<0|\alpha_r (x)
\beta_{r^\prime} (y)|0\right> - 
\theta (y^+-x^+)\left<0|\beta_{r^\prime} (y)
\alpha_r (x)|0\right>\ ;
\eqno(3.16)
$$
then the propagator (3.15) can be cast into a matrix form: namely,
$$
iS^+ (x-y) =\left|\matrix{S^{+\psi\chi^\dagger}\sigma^1 & 
S^{+\psi\psi^\dagger}\sigma^1\cr
S^{+\chi\chi^\dagger}\sigma^1 & S^{+\chi\psi^\dagger}\sigma^1\cr}\right|\ .
\eqno(3.17)
$$

The only independent light-front pairing turns out to be
$$
S^{+\psi\psi^\dagger}(x-y)=\tau^2\sqrt2\partial_- D(x-y;m)\ ,
\eqno(3.18)
$$
where
$$
D(x-y;m)=\int {d^4 k\over (2\pi)^4}{i\over k^2-m^2+i\epsilon}e^{ik(x-y)}
\eqno(3.19)
$$
is the free propagator of the massive real scalar field. Then, from the
constraints (3.8) we eventually obtain
$$
\eqalignno{
S^{+\psi\chi^\dagger}(x-y) & = \left(i\tau^{\alpha\dagger}\partial_\alpha
+m\sigma^1\right)D(x-y;m)\ , & (3.20a)\cr
S^{+\chi\psi^\dagger}(x-y) & = \left(i\tau^{\alpha}\partial_\alpha
+m\sigma^1\right)D(x-y;m)\ , & (3.20b)\cr
S^{+\chi\chi^\dagger}(x-y) & = \tau^2\sqrt2{\partial_\perp^2 - m^2\over 
2\partial_-}D(x-y;m)\cr
& = \tau^2\sqrt2\partial_+D(x-y;m)+i\tau^2{1\over \sqrt2\partial_-}
\delta^{(4)}(x-y)\ .
& (3.20c)\cr}
$$
As a consequence, from Eq.~(3.17) and taking Eq.~(3.1) into account,
the free fermion light-front propagator can be written in the form
$$
iS^+ (x-y) = \left(i\gamma^\mu\partial_\mu + m\right)D(x-y;m)
- {\gamma^+\over 2\partial_-}\delta^{(4)}(x-y)\ ,
\eqno(3.21)
$$
where the first term in the RHS is the usual covariant fermion propagator
$$
S^{{\rm cov}}(x-y)=\int {d^4 k\over (2\pi)^4}
{m-\gamma^\mu k_\mu\over k^2-m^2+i\epsilon}e^{ik(x-y)}\ ,
\eqno(3.22)
$$
whilst the second one is the so called "instantaneous" or "contact" term, which
is generated by the propagation along the light-cone generating lines.
The role of those term will be further elucidated in the next sections; in
particular, it will be clear that there is no need to specify any prescription
to define the light-front-space anti-derivative $\partial_-^{-1}$
which appears in eq.~(3.21).
\bigskip
\noindent
{\bf 4.\ Light-front Q.E.D. in the light-cone temporal gauge.}
\bigskip
We are now ready to discuss the main subject, {\it i.e.} the perturbative
light-front formulation of spinor Q.E.D., in which the LCC $x^+$ plays the
role of evolution parameter, within the light-cone gauge choice $A_+=0$.
Owing to this pattern (the controvariant LCC $x^+$ just corresponds to
the covariant component $A_+$ of the abelian vector potential), this 
formulation will be naturally referred to as light-front Q.E.D. in the
light-cone temporal gauge.

The starting point is obviously the lagrangean density
$$
{\cal L}= -{1\over 4}F_{\mu\nu}F^{\mu\nu}-\Lambda A_+
+\bar \Psi\left(i\gamma^\mu\partial_\mu -m\right)\Psi
+ e A_\mu\bar \Psi\gamma^\mu\Psi\ ,
\eqno(4.1)
$$
which can be rewritten, using the notations of the previous section, in
the form
$$
\eqalign{
{\cal L} & = -{1\over 4}F_{\mu\nu}F^{\mu\nu}-\Lambda A_+\cr
& + \psi^\dagger i\sqrt{2}\partial_+\psi +
\chi^\dagger i\sqrt{2}\partial_-\chi +
\psi^\dagger\left(i\tau^{\alpha\dagger}\partial_\alpha -m\sigma^1\right)\chi
 +\chi^\dagger\left(i\tau^{\alpha}\partial_\alpha -m\sigma^1\right)\psi \cr
& + eA_+\sqrt2 \psi^\dagger\psi + e 
A_\alpha\left(\psi^\dagger\tau^{\alpha\dagger}
\chi +\chi^\dagger\tau^\alpha\psi\right) + e A_-\sqrt2 \chi^\dagger\chi\ .\cr}
\eqno(4.2)
$$

As the interaction does not contain derivative couplings, the definitions of the
canonical momenta do not change with respect to the free case: then we have,
$$
\eqalignno{
\pi^-  
      &  = F_{+-}\ , & (4.3a)\cr
\pi^\alpha 
      &  = F_{-\alpha}\ , & (4.3b)\cr
\pi^+
      &  = 0\ , & (4.3c)\cr
\pi^\Lambda
      &  = 0\ , & (4.3d)\cr
\pi^\psi & = -i\sqrt{2}\psi^\dagger\ , & (4.3e)\cr
\pi^{\psi^\dagger} & = 0\ , & (4.3f)\cr
\pi^{\chi} & = 0\ , & (4.3g)\cr
\pi^{\chi^\dagger} & = 0\ . & (4.3h)\cr}
$$
where, again, (4.3b-e-f) are primary second class constraints whilst 
the remaining ones, but eq.~(4.3a), are primary first class. 
The canonical hamiltonian
reads
$$
\eqalign{
H & =\int d^3 x\left\{{1\over 2}\left(\pi^-\right)^2 + {1\over 4}
F_{\alpha\beta}F_{\alpha\beta} - A_+\left(\partial_\alpha\pi^\alpha
+ \partial_-\pi^- - \Lambda\right)\right.\cr
& -\chi^\dagger i\sqrt{2}\partial_-\chi -
\psi^\dagger\left(i\tau^{\alpha\dagger}\partial_\alpha -m\sigma^1\right)\chi
 -\chi^\dagger\left(i\tau^{\alpha}\partial_\alpha -m\sigma^1\right)\psi\cr
& \left.- eA_+\sqrt2 \psi^\dagger\psi - 
e A_\alpha\left(\psi^\dagger\tau^{\alpha\dagger}
\chi + \chi^\dagger\tau^\alpha\psi\right) + e A_-\sqrt2 \chi^\dagger\chi
\right\}\cr}
\eqno(4.4)
$$
and from the light-front temporal consistency of the primary first class
constraints the following secondary constraints arise: namely,
$$
A_+=0\ ,
\eqno (4.5a)
$$
$$
\partial_\alpha\pi^\alpha +\partial_-\pi^- - \Lambda 
- e\sqrt2\psi^\dagger\psi = 0\ ,
\eqno(4.5b)
$$
$$
\eqalignno{
& i\sqrt{2} D^*_-\chi^\dagger + i D^*_\alpha\psi^\dagger
\tau^{\alpha\dagger} + m\psi^\dagger\sigma^1 = 0\ , & (4.5c)\cr
& i\sqrt{2} D_-\chi +\left(i\tau^\alpha D_\alpha 
- m\sigma^1\right)\psi = 0\ , & (4.5d)\cr}
$$
where, as usual, we have set $D_\mu\equiv \partial_\mu - ieA_\mu$.

The whole set of primary and secondary constraints is now second class
and we can proceed to the calculation of graded Dirac's brackets.
To this aim, however, it is better to make a preliminary observation.
From the constraint equations (4.5c-d) it is apparent that, if we 
want to express the two-components spinors $\chi$ and $\chi^\dagger$
as functionals of the independent ones $\psi$ and $\psi^\dagger$,
we have to invert the differential operator
$D_- =\partial_- - ie A_-$. In the present context the corresponding
Green's function will be understood as a formal series:
namely,
$$
{1\over D_-}\equiv {1\over \partial_-}\sum_{n=0}^\infty
\left(ieA_-{1\over \partial_-}\right)^n\ ,
\eqno(4.6)
$$
where each anti-derivative acts upon all the factors on its right.

As it will be clear later on, we remark that it is neither necessary
nor convenient to specify any kind of prescription, in order to
properly define the anti-derivative itself. Furthermore, it is unavoidable that
the Dirac's hamiltonian, in which all the constraints are solved in
terms of the independent fields, would result into a formal (infinite)
power series of the dimensionless electric charge $e$.

Let us turn now to the calculation of the graded Dirac's brackets.
As the actual inversion of the constraints matrix is a little bit complicated
in the present case, it is convenient to operate iteratively and compute
some sequences of preliminary brackets (eventually four sequences).
After taking
$$
\xi_1\equiv A_1 ,\ \xi_2\equiv A_2 ,\ \xi_3\equiv A_- ,\
\xi_4\equiv \pi^- ,\ \xi_5\equiv \psi ,\ \xi_6\equiv \psi^\dagger ,
\eqno(4.7)
$$
as independent fields, a straightforward although very tedious calculation
leads to the following result: namely,
$$
\Xi_{ab}({\bf x},{\bf y})=\left|\matrix{
-{\bf 1}/2\partial_- & 0 & 0 & \partial_1/2\partial_- & 0 & 0\cr
0 & -{\bf 1}/2\partial_- & 0 & \partial_2/2\partial_- & 0 & 0\cr
0 & 0 & 0 & {\bf 1} & 0 & 0\cr
-\partial_1/2\partial_- & -\partial_2/2\partial_- & 
-{\bf 1} & \partial_\perp^2/2\partial_- & 0 & 0\cr
0 & 0 & 0 & 0 & 0 & -i/\sqrt2\cr
0 & 0 & 0 & 0 & -i/\sqrt2 & 0\cr}\right|\ ,
\eqno(4.8)
$$
where, once again, we have denoted the Dirac's brackets matrix as
$$
\Xi_{ab}({\bf x},{\bf y})\equiv
\left.\left\{\xi_a (x),\xi_b (y)\right\}_D\right|_{x^+=y^+}\ ,\quad
a,b=1,\ldots ,6\ .
$$

It is important to realize that the set of the independent interacting
fields $\xi_a (x) ,\ a=1,\ldots ,6$, do obey the very same algebra as the
corresponding independent free fields, notwithstanding the fact that the
secondary constraints are quite different in the two cases.
This feature, as we shall see in the sequel, is of crucial importance
in setting up the perturbation theory. Moreover, it has to be gathered 
that the above property does not hold in general for an arbitrary
constrained system, but it depends, in the present case, upon a
clever choice of the independent fields.

Finally, after solving the secondary constraints in terms of the 
independent fields $\xi_a (x) ,\ a=1,\ldots ,6$, the hamiltonian
(4.4) takes its Dirac's form which becomes
$$
\eqalign{
H_D & =\int d^3 x\left\{{1\over 2}\left(\pi^-\right)^2 + {1\over 4}
F_{\alpha\beta}F_{\alpha\beta}\right.\cr
& \left.- \left(i\partial_\alpha\psi^\dagger\tau^{\alpha\dagger}
+ m\psi^\dagger\sigma^1 - eA_\alpha\psi^\dagger\tau^{\alpha\dagger}\right)
{1\over i\sqrt2 D_-}
\left(i\tau^{\alpha}\partial_\alpha\psi
- m\sigma^1\psi + eA_\alpha\tau^\alpha\psi\right)\right\}\ ,\cr}
\eqno(4.9)
$$
which is the starting point to develop perturbation theory as we discuss
in the next section.
\bigskip
\noindent
{\bf 5.\ Perturbation theory.}
\bigskip
In order to separate the interaction hamiltonian in a constrained system,
one has to be very careful in the choice of the independent canonical
variables: as a matter of fact, the basic criterion to select the latter
ones is eventually dictated by the structure of the Dirac's brackets of the
interacting theory.

On the one hand, after choosing $\xi_a (x) ,\ a=1,\ldots ,6$ as independent
fields, we see that the first line of the RHS of Eq.~(4.9) does not contain
the coupling constant $e$ and, consequently, does not contribute to the
interaction hamiltonian. On the other hand, had we chosen as independent
fields the set $A_1,\ A_2,\ A_-,\ \Lambda,\ \psi ,\ \psi^\dagger$, which
is a perfectly legitimate choice, then, after solving $\pi^-$ as a functional
of the above variables, we find that the first line in the RHS of Eq.~(4.9)
does indeed contribute to the interaction hamiltonian through 
the two terms:
$$
-e\sqrt2\left\{-\partial_\alpha A_\alpha +
{1\over \partial_-}\left(\partial_\perp^2 A_- +\Lambda\right)\right\}
{1\over \partial_-}\left(\psi^\dagger \psi\right)
+e^2\left\{{1\over \partial_-}\left(\psi^\dagger\psi\right)\right\}^2\ ,
$$
whence, thereby, a quite different kind of perturbation theory does follow.

In view of the above remark, one could be eventually led to the conclusion
that perturbation theory for constrained systems is not univocally determined,
owing to the fact that it depends upon the specific choice of the independent
fields, in terms of which the constraints are solved.
Actually, this apparent ambiguity is not there. As a matter of fact, we
recall that perturbation theory stems from the assumption of the existence, 
at least formally, of the so called evolution operator, which implements
the time-dependent unitary transformation relating the interacting to the free 
fields - see, for instance, [12].

On the other hand, we know that a unitary operator is such to preserve the
canonical equal time field algebra. This means that, in the case of 
constrained systems, the suitable 
independent interacting fields must fulfil the very
same equal time operator algebra as the corresponding free fields do. 
In terms
of those, and only those, independent interacting fields the interaction
hamiltonian has to be expressed and perturbation theory
will be safely and consistently developed .

From the constraints (4.3b), (4.5b) and the Dirac's brackets (4.8), it is
an easy exercise to show that
$$
\left.\left\{\Lambda (x),\psi (y)\right\}_D\right|_{x^+ = y^+} =
i e\psi (x)\delta^{(3)}({\bf x}-{\bf y})\ ,
\eqno(5.1)
$$
whereas, in the free field case, the corresponding Dirac's bracket vanishes.
As a consequence, the construction of the interacting hamiltonian as a
functional of the fields $A_1,\ A_2,\ A_-,\ \Lambda,\ \psi ,\ \psi^\dagger$
does not make sense in order to set up perturbation theory.
The interaction hamiltonian is expressed in terms of the set of 
independent fields $\xi_a (x) ,\ a=1,\ldots ,6$, whose Dirac's brackets (4.8)
do not depend upon the electric charge $e$, what makes it now 
clear why the above algebra (4.8) has been precisely put forward. 

We now consider the second line of the hamiltonian (4.9). As all 
the field operators in the interaction picture evolve according to free 
equations of motion, it is convenient to replace with $\chi$ and $\chi^\dagger$
those linear combinations of the fields $\psi$ and $\psi^\dagger$, which 
coincide with the solutions of the free constraint equations (3.8a-b).
After this, we can rewrite the hamiltonian (4.9) in the form:
$$
\eqalign{
H_D & =\int d^3 x\left\{{1\over 2}\left(\pi^-\right)^2 + {1\over 4}
F_{\alpha\beta}F_{\alpha\beta}\right.\cr
& +\left.\left(i\sqrt{2}\partial_-\chi^\dagger
+ eA_\alpha\psi^\dagger\tau^{\alpha\dagger}\right)
{1\over i\sqrt2 D_-}
\left(-i\sqrt{2}\partial_-\chi
+ eA_\alpha\tau^\alpha\psi\right)\right\}\ .\cr}
\eqno(5.2)
$$

If we now perform, within the second line of the above equation, 
the following replacements: namely,
$$
\eqalignno{
\chi & \longmapsto {1\over 2}\gamma^+\gamma^-\Psi\ , & (5.3a) \cr
\tau^\alpha\psi & \longmapsto {1\over \sqrt{2}}\gamma^+\gamma^\alpha\Psi\ ,
& (5.3b) \cr
\chi^\dagger & \longmapsto {1\over \sqrt{2}}\bar\Psi\gamma^-\ , & (5.3c) \cr
\psi^\dagger\tau^{\alpha\dagger} & \longmapsto {1\over 2}\bar\Psi
\gamma^\alpha\gamma^+\gamma^-\ , & (5.3d) \cr}
$$
we eventually obtain
$$
H_D  =\int d^3 x\left\{{1\over 2}\left(\pi^-\right)^2 + {1\over 4}
F_{\alpha\beta}F_{\alpha\beta} +\bar\Psi\gamma^- i\partial_-\Psi -
eA_\mu\bar\Psi\gamma^\mu\Psi +
{e^2\over 2}A_\mu\bar\Psi\gamma^\mu{1\over iD_-}A_\nu\gamma^\nu\Psi\right\}\ .
\eqno(5.4)
$$

It is evident, from the above final form of the Dirac's hamiltonian, that the 
interaction hamiltonian density, upon which perturbation theory is set, reads
$$
{\cal H}_{{\rm int}} = -eA_\mu\bar\Psi\gamma^\mu\Psi +
{e^2\over 2}A_\mu\bar\Psi\gamma^\mu{\gamma^+\over iD_-}A_\nu\gamma^\nu\Psi\ .
\eqno(5.5)
$$
It is now apparent that, besides the usual covariant vertex of Q.E.D., 
we have to 
consider, taking the formal definition (4.6) into account, an infinite number
of non-covariant vertices. On the other hand, we have seen that also the free
Dirac's propagator (3.21) exhibits a non-covariant term besides the usual one.
What happens, as we shall here explicitely show up to the one loop order, 
is that
in dimensionally regularized truncated Green's functions all 
those non-covariant terms cancel, leaving us with the very same renormalizable 
one loop structures, as found in the standard STC framework [8].

To this aim, let us first obtain the Feynman's rules. From the definition (4.6)
together with the identity
$$
{\gamma^+\over 2}A_-  =  {\gamma^+\over 2}A_\nu\gamma^\nu {\gamma^+\over 2}\ ,
\eqno(5.6)
$$
we can formally expand the interaction hamiltonian density as
$$
\eqalign{
i{\cal H}_{{\rm int}} & = ie A_\mu\bar\Psi \gamma^\mu\Psi\cr
& -ieA_\mu\bar\Psi\gamma^\mu{\gamma^+\over 2\partial_-}ieA_\nu\gamma^\nu \Psi\cr
& -ieA_\mu \bar\Psi \gamma^\mu{\gamma^+\over 2\partial_-}
ieA_\rho\gamma^\rho{\gamma^+\over 2\partial_-}ieA_\nu\gamma^\nu \Psi\cr
& -ieA_\mu\bar\Psi\gamma^\mu{\gamma^+\over 2\partial_-}
         ieA_\rho\gamma^\rho{\gamma^+\over 2\partial_-}
         ieA_\sigma\gamma^\sigma{\gamma^+\over 2\partial_-}
         ieA_\nu\gamma^\nu\Psi +\ldots \ ,\cr}
\eqno(5.7)
$$
where the anti-derivatives (integral operators) act upon all 
the factors on their right.

From eq.s~(2.35), (3.21) and (5.7), we get the Feynman's rules listed in 
Fig.2.
\vfill\eject
\midinsert
\vskip 20 truecm
\includegraphics{figura2bis.eps} 
\endinsert
\vfill\eject
Using these rules, it is not difficult to check graphically that in the one
loop truncated Green's functions, but photon self-energy diagram, all the
non-covariant terms cancel algebraically.

For instance, taking two covariant vertices (Fig.2e) and a second order
non-covariant one (Fig.2f), we reconstruct the full one loop electron
self-energy (see Fig.3a), which, after the removal of the external legs, 
turns out to be the correct renormalizable one of the standard STC approach.
\midinsert
\vskip 5 truecm
\includegraphics{figura3a.eps}
\endinsert
Moreover, the one loop renormalizable electron-positron-photon proper vertex 
can be reconstructed (see Fig.3b) taking the covariant vertices of Fig.2e 
as well as first and second order non-covariant vertices of Fig.s~2f-g 
into account.
\midinsert
\vskip 6 truecm
\includegraphics{figura3b.eps}
\endinsert
Let us come now to the photon one loop self-energy of Fig.3c.
\vfill\eject 
\midinsert
\vskip 8 truecm
\includegraphics{figura3c.eps}
\endinsert

After 
summation of the relevant vertices, we see that, beside the correct
standard diagram, a further non-covariant graph is there, whose corresponding
integral (in $2\omega$ space-time dimensions) is provided by:
$$
I^{\rho\sigma}(p_-)=
(ie)^2\int {d^{2\omega}l\over (2\pi)^{2\omega}}Tr\left\{\gamma^\rho{\gamma^+\over
2i(l_-+p_-)}\gamma^\sigma{\gamma^+\over 2il_-}\right\}\ .
\eqno(5.8)
$$
However, since
$$
\int d^{2\omega}l =\int dl_+\int dl_-\int d^{2\omega-2}l_\perp \ ,
$$
we immediately see that integration over transverse momenta in (5.8) gives a
vanishing result. This is the only point, up to the one loop approximation,
in which the cancellation of non-covariant vertices does not take place
algebraically but involves a further analytic tool. Owing to the above
cancellation mechanisms, either algebraic or due to dimensional regularization
of integrals over transverse momenta, it becomes clear why it is immaterial
to specify any prescrition to understand non-covariant denominators in
fermions propagators as well as in the interaction vertices, at least in
perturbation theory.

To sum up, we have shown that, concerning one loop dimensionally regularized
truncated Green's functions, the light-front formulation of Q.E.D. in the 
light-cone temporal gauge actually reproduces the very same result as in the 
standard STC renormalizable and (perturbatively) unitary 
approach [8], 
in which non-covariant singularities are regulated by means of the
ML prescription.
\bigskip
\noindent
{\bf 6.\ Conclusion.}
\medskip
A consistent light-front formulation of perturbative Q.E.D. has been 
worked out in the
light-cone gauge $A_+=0$, in which the LCC $x^+$ plays the role of the
evolution parameter. Owing to this, it is natural, by analogy with the 
ordinary STC formulation, to refer our choice as to "temporal" 
light-cone gauge, alternative to the original 
"axial" choice $A_-=0$.
By consistent, we understand that the quantization scheme here developed
reproduces, at least up to the one loop order, the same off-shell amplitudes 
as computed from the conventional correct approach in usual STC [7],~[8],
which embodies the ML prescription to define the spurious non-covariant
singularities of the free photon propagator.

This result is non-trivial and, in turn, also rather surprising.
As a matter of fact, it has been thoroughly unravelled [14] that in the 
quantization of gauge theories in ordinary STC, the use of the temporal 
(or Weyl) subsidiary condition $A_0=0$ is undoubtely much more troublesome 
than the axial one $A_3=0$,
which is in turn also affected by subtle mathematical pathologies [15].
Eventually, in spite of the huge number of attempts
and efforts, the problem of setting up a fully consistent perturbation   
theory in the temporal gauge is still to be solved.

On the contrary, within the light-front perturbative formulation, 
the "temporal" gauge
choice $A_+=0$ appears to be the safe one, which naturally leads to the
ML prescrition and thereby to the equivalence with the convention approach
in STC, whilst the "spatial" choice $A_-=0$ drives to inconsistency [5].

A further comment is deserved to the gauge invariace of the regularization 
methods in perturbation theory. It clearly appears that, in the present
context, the use of dimensional regularization is crucial, in order to
provide an infinite set of diagrams cancellation, in the absence of which
gauge invariance of Q.E.D. would be lost. Things are not so lucky for
cut-off or Pauli-Villars regularizations, which, thereof, turn out to be
quite inconvenient within the perturbative light-front approach.

It should be noticed that the presence of an infinite number of non-covariant
vertices, switching on order-by-order in light-front perturbation theory,
closely figures the structure of counterterms for
the 1PI-vertices in the standard STC approach
to the light-cone gauges [8]. This feature is probably connected to the
specific properties of the ML propagator, 
{\it i.e.} to the kind of structures it
generates after loop integrations.

Although graphically transparent, 
a formal general proof - which is basically by induction -
that the cancellation mechanism for non-covariant terms persists,
to all order in perturbation theory, will be presented in a forthcoming
paper, together with the generalization of the present
treatment to the non-abelian case. 

\bigskip
\noindent
{\bf Acknowledgments.}
\medskip
One of us (R. S.) warmly thanks G. McCartor and D. G. Robertson for
quite useful conversations, correspondence and discussions and
is grateful to G. Nardelli for reading the manuscript.
This work has been supported by a MURST grant, quota 40\%.

\bigskip
\noindent
{\bf References}
\medskip
\item{[1]}\ B. D. Jones, R. J. Perry : Phys. Rev. {\bf D55} (1997) 7715;
            M. M. Brisudova, R. J. Perry : Phys. Rev. {\bf D54} (1996) 6453.

\item{[2]}\ B. D. Jones : {\tt hep-th/9703106} ; 
            B. D. Jones, R. J. Perry, S. D. Glazek : Phys. Rev. {\bf D55} 
            (1997) 6561; 
            M. Krautg\"artner, H. C. Pauli, F. W\"olz : Phys. Rev. {\bf D45}
            (1992) 3755; 
            J. R. Hiller : "{\it Theory of Hadrons and Light-Front
            QCD}", World Scientific, Singapore (1995) 277-281.

\item{[3]}\ A. C. Kalloniatis, D. G. Robertson : Phys. Rev. {\bf D50} (1994)
            5262; 
            D. G. Robertson : "{\it Theory of Hadrons and Light-Front
            QCD}", World Scientific, Singapore (1995) 111-116; 
            A. B. Bylev, V. A. Franke, E. V. Prokhvatilov : 
            {\tt hep-th/9711022}; 
            K. Yamawaki : {\tt hep-th/9707141}.

\item{[4]}\ J. B. Kogut, D. E. Soper : Phys. Rev. {\bf D1} (1970) 2901;
            F. R\"ohrlich : Acta Phys. Austriaca, Suppl. VIII (1971) 2777;
            R. A. Neville, F. R\"ohrlich : Phys. Rev. {\bf D3} (1971) 1692;
            J. M. Cornwall, R. Jackiw : Phys. Rev. {\bf D4} (1971) 367;
            E. Tomboulis : Phys. Rev. {\bf D8} (1973) 2736;
            J. H. Ten Eyck, F. R\"ohrlich : Phys. Rev. {\bf D9} (1974) 436.

\item{[5]}\ D. M. Capper, J. J. Dulwich, M. J. Litvak : Nucl. Phys. {\bf 241B}
            (1984) 463.

\item{[6]}\ S. Mandelstam : Nucl. Phys. {\bf 213B} (1983) 149;
            G. Leibbrandt : Phys. Rev. {\bf D29} (1984) 1699.

\item{[7]}\ A. Bassetto, M. Dalbosco, I. Lazzizzera, R. Soldati : Phys. Rev.
            {\bf D31} (1985) 2012.

\item{[8]}\ A. Bassetto, M. Dalbosco, R. Soldati : Phys. Rev. {\bf D36} (1987)
            3138; 
            C. Becchi : "{\it Physical and Nonstandard Gauges}",
            Springer-Verlag, Berlin (1990) 177-184.
\item{[9]}\ G. McCartor, D. G. Robertson : Z. Phys. {\bf C62} (1994) 349,
            {\bf C68} (1995) 345.  
    
\item{[10]}\ G. McCartor : Z. Phys. {\bf C41} (1988) 271.

\item{[11]}\ P. A. M. Dirac : "{\it Lectures on Quantum Mechanics}",
             Belfer Graduate School of Science - Yeshiva University,
             Academic Press, New York (1964);
             E. C. G. Sudarshan, M. Mukunda : "{\it Classical
             Dynamics - A Modern Perspective}", Wyley Interscience,
             New York (1974).
 
\item{[12]}\ R. Casalbuoni : Nuovo Cim. {\bf 33A} (1976) 115.

\item{[13]}\ C. Itzykson, J. B. Zuber : "{\it Quantum Field Theory}",
             McGraw-Hill Int. Ed., Singapore (1985).

\item{[14]}\ A. Bassetto, G. Nardelli, R. Soldati : "{\it Yang-Mills Theories 
             in Algebraic Non-Covariant Gauges}", World Scientific, Singapore 
             (1991).

\item{[15]}\ G. Nardelli, R. Soldati : Int. Jour. Mod. Phys. {\bf 5} (1990)
             3171.

\vfill\eject\end

%Dear Dr. D. Nordstrom,

 %    herewith you will find a new version of our manuscript DM6437

%{\it
%"Consistent perturbative light-front formulation of quantum electrodynamics"
%}

%we would like to resubmit to Phys. Rev. D for publication.
 %    We have attempted to fulfil, as close as possible, all the 
%recommendations and criticisms raised by the referee, as summarized below.
 %    We hope that, in the present form, our paper could be accepted for
%publication on Phys. Rev. D.

 %    With my best regards

%Roberto Soldati
\vfill\eject
%%%%%%%%%%%%%%%%%%%%%%%%%%%%%%%%%%%%%%%%%%%%%%%%%%%%%%%%%%%%%%%%%%%%%%%%%%%%
%\noindent
%Response to the Referee's report.
\bigskip
i)  We agree with the Referee that the previous title exhibited an important
omission, the key word "perturbative": accordingly, we have changed the title
and inserted the key word at several places within the manuscript.
\medskip
ii) It was certainly far from our aims to adopt "strong wording against"
and/or to make "unfair criticisms" to the papers by G. McCartor and D. G.
Robertson: on the contrary, we were very well aware of the important 
novelties and hints contained in those papers and we honestly tried to
emphasize their relevance with respect to our own developments.
Moreover, we are good friends and have discussed the matter in several
occasions (we added something in the acknowledgments, as it was due): 
as far as we know, they are perfectly aware of our results
and did not find anything offending in our writing.  
Nonetheless, the first one of our [9], namely Z. Phys. {\bf C62} (1994) 349,
actually contains an incorrect statement, {\it i.e.}, 
it is not true that eq. (2.41)
(here there is light-front-time ordering, not ordinary time as alluded
by the Referee!) together with expansions (2.42-43) give rise to the
{\it "ML form of $D^{--}$"}. 
So, apart from the practical purposes, as correctly
emphasized by the Referee, we think better to alert the reader that,
strictly speaking, our quantization scheme actually leads to the ML form 
of the light-front vector propagator, whereas the approach of [9] does not.
Of course, the latter might be quite consistent, we do not write the opposite,
but it has to be proved and, in any case, the ML form of the light-front
propagator does not follow thereoff. But that's all, we do not honestly
find anything unfair in this technical remark: anyway, we have further 
smoothed the sentence 
       {\it "One can easily convince himself ...." }
according to the Referee's suggestion.
\medskip
iii) In our paper we restrict ourselves to the perturbative domain and to
the abelian gauge group. Within those bounds, we have found a light-front
formulation which turns out to be equivalent to the ordinary time approach
of our Ref.s [7-8]. We find this result remarkable. However, we fully agree
with the Referee that many important and unanswered questions are there
concerning, in particular, non-abelian and non-perturbative issues.
What we think is that any well established light-front formulation of
gauge theories must, at least, correctly reproduce the well known
order-by-order perturbative results: this is a kind of "minimal test"
in order to be trusted. But if and how our abelian perturbative analysis
could go over deep and hard non-abelian and non-perturbative topics, we
do not honestly know at this stage. We prefer to avoid any comments
to that concern and stress everywhere else (starting from the very title)
that we are within the abelian perturbative domain and nothing more: we
would like to avoid any "background fanfare" as much as possible.

     We hope that, in so doing, all the Referee's recommendations and
criticisms are fully taken into account and our new version of the
manuscript is suitable for publication on Phys. Rev. D.
\vfill\eject

%%%%%%%%%%%%%%%%%%%%%%%%%%%%%%%%%%%%%%%%%%%%%%%%%%%%%%%%%%%%%%%%%%%%%%%%%%%%